\documentclass[aps,prl,reprint,amssymb,showpacs,superscriptaddress,notitlepage]{revtex4-1}
\usepackage{bm}
\usepackage{graphicx}
\usepackage{dcolumn}
\usepackage{braket}
\usepackage{natbib}
\usepackage{amsmath}
\usepackage{color}
\usepackage{ulem}
\usepackage{subcaption} 
\usepackage{siunitx}
\usepackage{adjustbox}
\definecolor{gray}{rgb}{0.7,0.7,0.7}
\definecolor{orange}{rgb}{1, 0.4, 0}
\definecolor{dgreen}{rgb}{0.0, 0.4, 0.0}
\definecolor{yblue}{rgb}{0.06, 0.3, 0.57}

\newcommand\ldsout{\bgroup\markoverwith{\textcolor{blue}{\rule[0.5ex]{2pt}{0.4pt}}}\ULon}

\usepackage{subfiles}
\begin{document}

\title{Bose-Einstein Condensation in Gap-Confined Exciton-Polariton States}

 \author{F. Riminucci}
 \affiliation{Molecular Foundry, Lawrence Berkeley National Laboratory, One Cyclotron Road, Berkeley, California 94720, USA}
 \author{A. Gianfrate}
 \affiliation{CNR Nanotec, Institute of Nanotechnology, via Monteroni, 73100, Lecce}
 \author{D. Nigro}
 \affiliation{Dipartimento di Fisica, Universit\`a di Pavia, via Bassi 6, Pavia (IT)}
 \author{V. Ardizzone}
 \affiliation{CNR Nanotec, Institute of Nanotechnology, via Monteroni, 73100, Lecce}
 \author{S. Dhuey}
 \affiliation{Molecular Foundry, Lawrence Berkeley National Laboratory, One Cyclotron Road, Berkeley, California 94720, USA}
 \author{L. Francaviglia}
 \affiliation{Molecular Foundry, Lawrence Berkeley National Laboratory, One Cyclotron Road, Berkeley, California 94720, USA}
 \author{K. Baldwin}
 \affiliation{PRISM, Princeton Institute for the Science and Technology of Materials, Princeton University, Princeton, New Jersey 08540, USA}
 \author{L. N. Pfeiffer}
 \affiliation{PRISM, Princeton Institute for the Science and Technology of Materials, Princeton University, Princeton, New Jersey 08540, USA}
 \author{D. Trypogeorgos}
 \affiliation{CNR Nanotec, Institute of Nanotechnology, via Monteroni, 73100, Lecce}
 \author{A. Schwartzberg}
 \affiliation{Molecular Foundry, Lawrence Berkeley National Laboratory, One Cyclotron Road, Berkeley, California 94720, USA}
 \author{D. Gerace}
 \affiliation{Dipartimento di Fisica, Universit\`a di Pavia, via Bassi 6, Pavia (IT)}
 \author{D. Sanvitto}
 \email{daniele.sanvitto@nanotec.cnr.it}
 \affiliation{CNR Nanotec, Institute of Nanotechnology, via Monteroni, 73100, Lecce}

\begin{abstract}
The development of patterned multi-quantum well heterostructures in GaAs/AlGaAs waveguides has recently allowed to achieve exciton-polariton condensation in a topologically protected bound state in the continuum (BIC). Remarkably, condensation occurred above a saddle point of the polariton dispersion. A rigorous analysis of the condensation phenomenon in these systems, as well as the role of the BIC, is still missing.
In the present Letter we theoretically and experimentally fill this gap, by showing that polariton confinement resulting from the negative effective mass and the photonic energy gap in the dispersion play a key role in enhancing the relaxation towards the condensed state. In fact, our results show that low-threshold polariton condensation is achieved within the effective trap created by the exciting laser spot regardless of whether the resulting confined mode is long-lived (polariton BIC) or short-lived (lossy mode). In both cases, the spatial quantization of the polariton condensate and the threshold differences associated to the corresponding state lifetime are measured and characterized. For a given negative mass, a slightly lower condensation threshold from the polariton BIC mode is found and associated to its suppressed radiative losses as compared to the lossy one.
\end{abstract}

\maketitle

\textit{Introduction}.
Exciton-polaritons are elementary excitations in solids that arise from the strong radiation-matter coupling between a confined photonic mode and an excitonic resonance. Their effects are enhanced when the excitonic field is confined in low-dimensional nanostructures, such as semiconductor quantum wells (QW). As hybrid radiation-matter bosonic-like excitations, they represent a powerful platform for either nonlinear optics applications\cite{Sturm2014All-opticalInterferometer,Suarez-Forero2021EnhancementInteractions,Zasedatelev2021Single-photonTemperature,Datta2022HighlyMoS2,Sanvitto2016}, including electrically pumped polariton lasers,\cite{Schneider2013} or studying fundamental physics phenomena such as out-of-thermal equilibrium Bose-Einstein condensation\cite{Kasprzak2006,Sun2017,Caputo2018TopologicalCondensates} and superfluidity\cite{Amo2009SuperfluidityMicrocavities,Lerario2017Room-temperatureCondensate}. 
Up to date, exciton-polariton condensation has been mostly limited to Distributed Bragg Reflector based planar microcavities with large vertical Q-factor. The high reflectivity of the Bragg mirrors is required to allow for polariton relaxation towards the dispersion energy minimum, where the critical particle density needed for condensation can be reached. For this reason, there has been great effort to laterally confine microcavity polaritons by suitably engineering the potential landscape to reduce their condensation threshold\cite{Yoon2022EnhancedPotentials,Sun2017,Schneider2017Exciton-polaritonEngineering,Ferrier2011InteractionsCondensates}.
Recently, polariton condensation has also been achieved in patterned waveguides without bottom and top Bragg mirrors, by exploiting QW exciton coupling to a long-lived photonic bound state in the continuum (BIC)\cite{Ardizzone2022PolaritonContinuum,Riminucci2022NanostructuredContinuum}. In these experiments, exciton-polaritons were shown to accumulate in the low energy BIC mode at normal incidence, corresponding to a saddle point of the energy dispersion, and separated by an energy gap from a radiative mode with a dispersion minimum (lossy branch)\cite{Hsu2016BoundContinuum,Kravtsov2020,Lu2020EngineeringPoints}. Even if this structure lacks Bragg mirrors, which greatly reduces the complexity of the required heterostructure to be grown, it still guarantees a low condensation threshold by suppressing the radiative losses through symmetry protection of the eigenmodes\cite{Zhen2014TopologicalContinuum}. However, a full theoretical account of the condensation mechanism, especially when considering the role of the negative effective mass around the saddle-like polariton dispersion, is still missing at time of writing, as well as a clear understanding of the actual relevance of the BIC lifetime in providing an efficient condensation mechanism.  

In this work, we unravel the critical role played by the negative mass and the energy gap between BIC and lossy modes in reaching the condensation onset in the saddle point observed in \cite{Ardizzone2022PolaritonContinuum}. We present both experimental evidence and the results of a theoretical modelling based on an extended out-of-equilibrium Gross-Pitaevskii formulation.
In particular, we emphasize the role of the blue-shifted negative-mass dispersion underneath the pump spot, which creates an optically-induced attractive potential. We then show that polariton condensation can be achieved in gap-confined spatially quantized eigenmodes, regardless of their BIC or lossy nature as induced by the symmetry properties of the lower-lying negative mass branch. Finally, we also highlight the differences between the two cases, showing how the polariton radiative losses still play a role in defining the condensation threshold. The presented theory and experimental results demonstrate the role of the negative mass and the energy gap around normal incidence to achieve polariton Bose-Einstein condensation in a waveguide-embedded multi-QW heterostructure.

\begin{figure*}
  \centering
  \begin{subfigure}{0.29\textwidth}
    \centering
    \includegraphics[width=\linewidth]{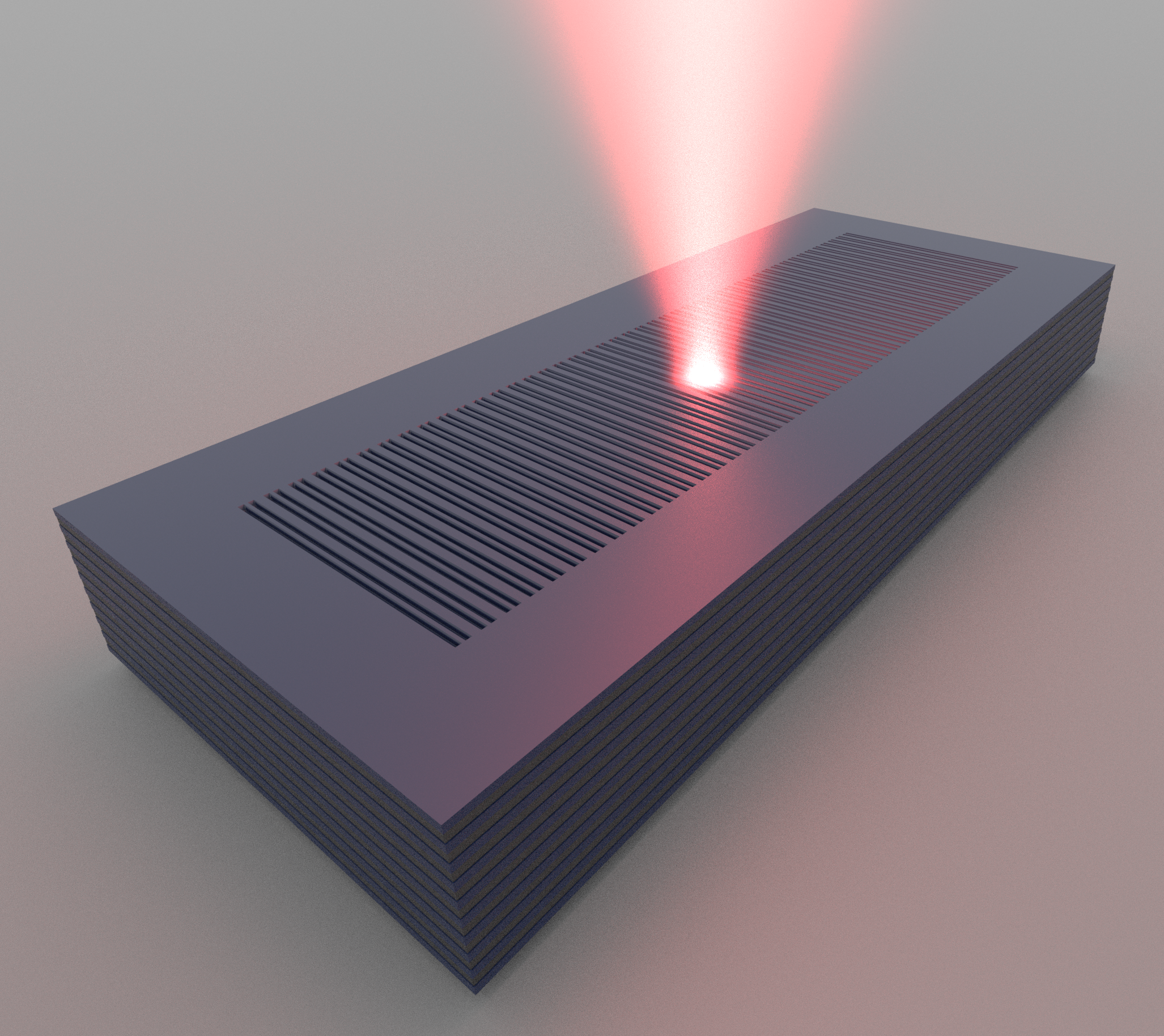}
    \put(-138,116){\textbf{a}}
  \end{subfigure}%
  \hspace{0.04\textwidth}%
  \begin{subfigure}{0.33\textwidth}
    \centering
    \includegraphics[width=\linewidth]{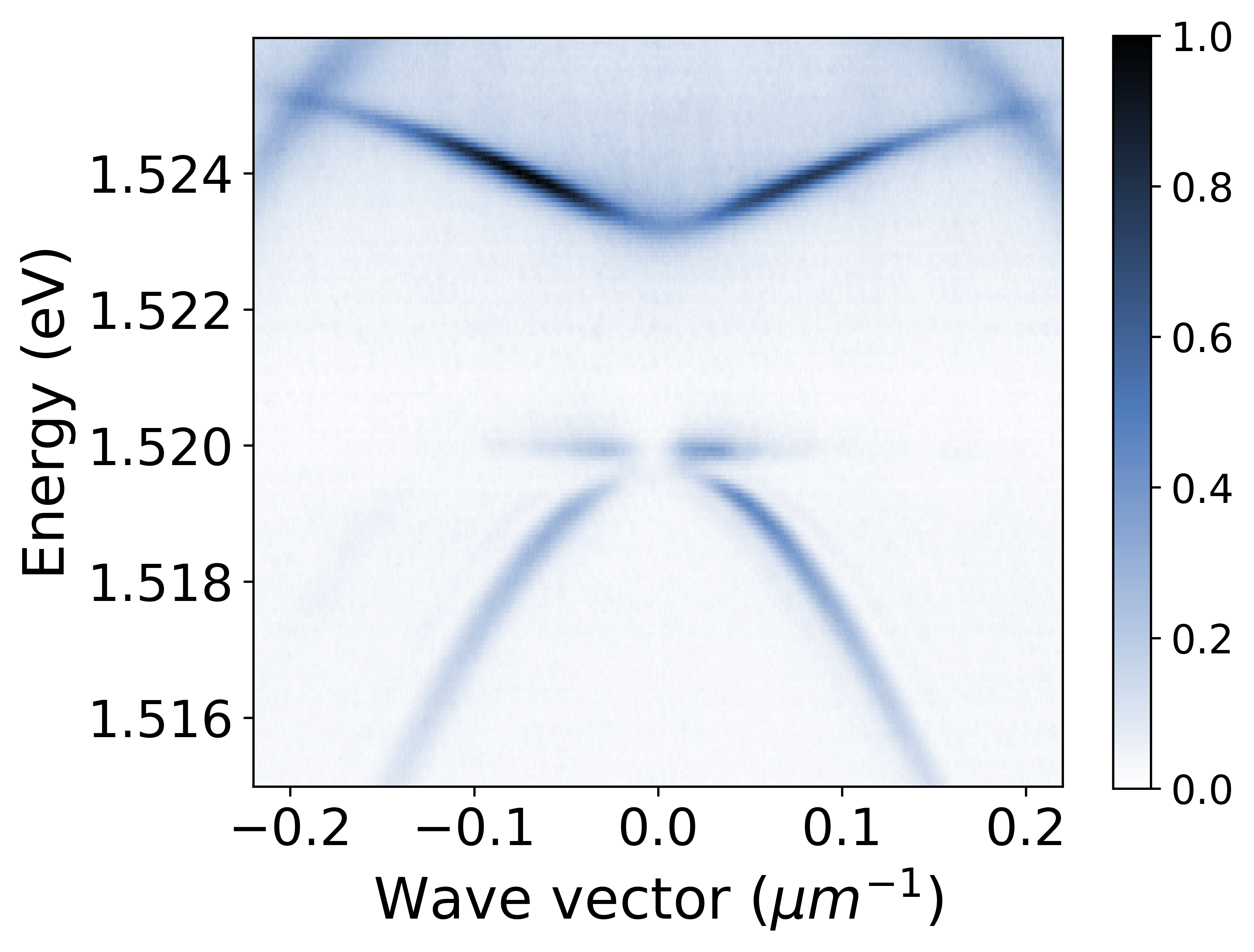}
    \put(-128,116){\textbf{b}}
  \end{subfigure}%
  \hspace{0.01\textwidth}%
  \begin{subfigure}{0.33\textwidth}
    \centering
    \includegraphics[width=\linewidth]{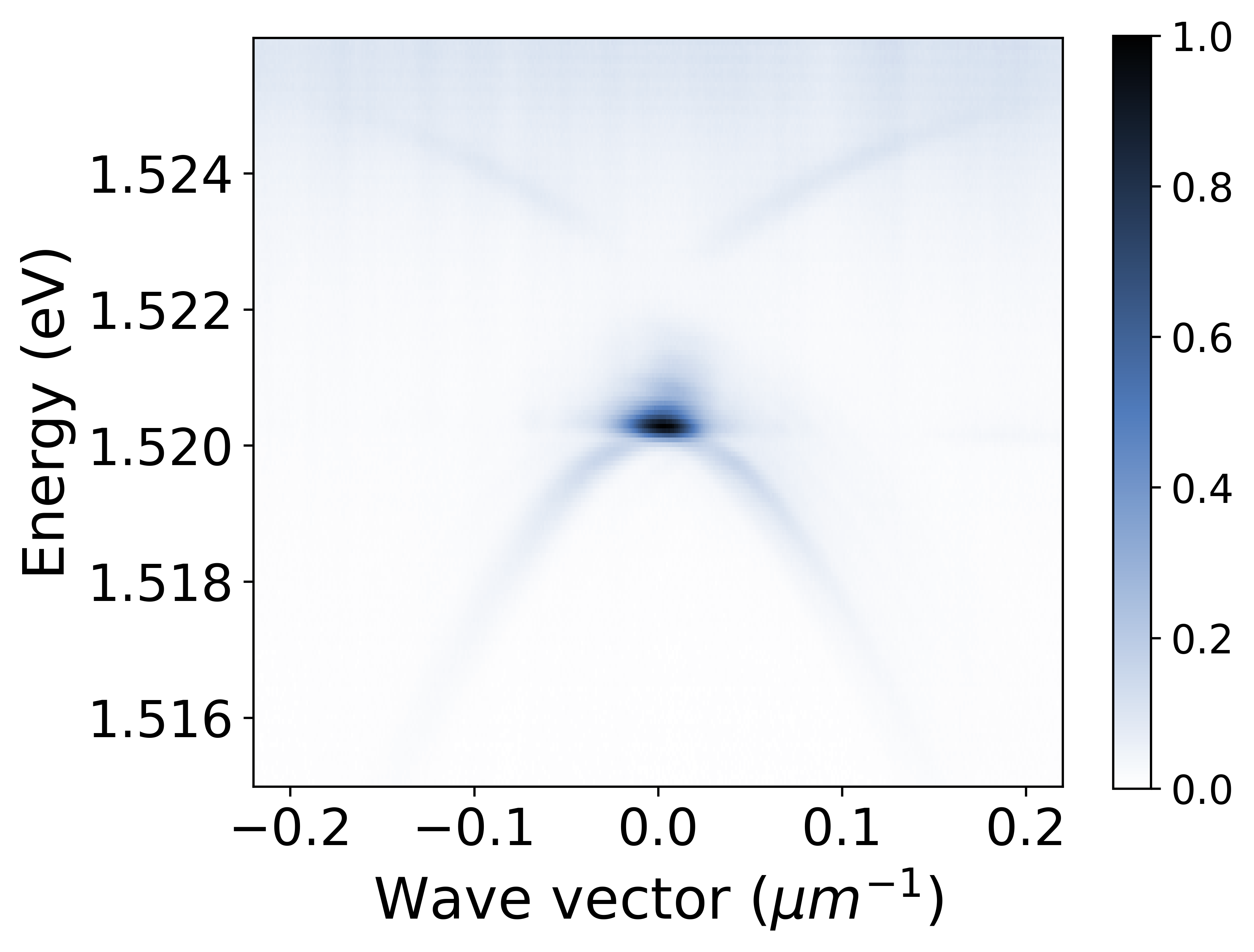}
    \put(-128,116){\textbf{c}}
  \end{subfigure}
  
  \medskip
  
  \begin{subfigure}{0.33\textwidth}
    \centering
    \includegraphics[width=\linewidth]{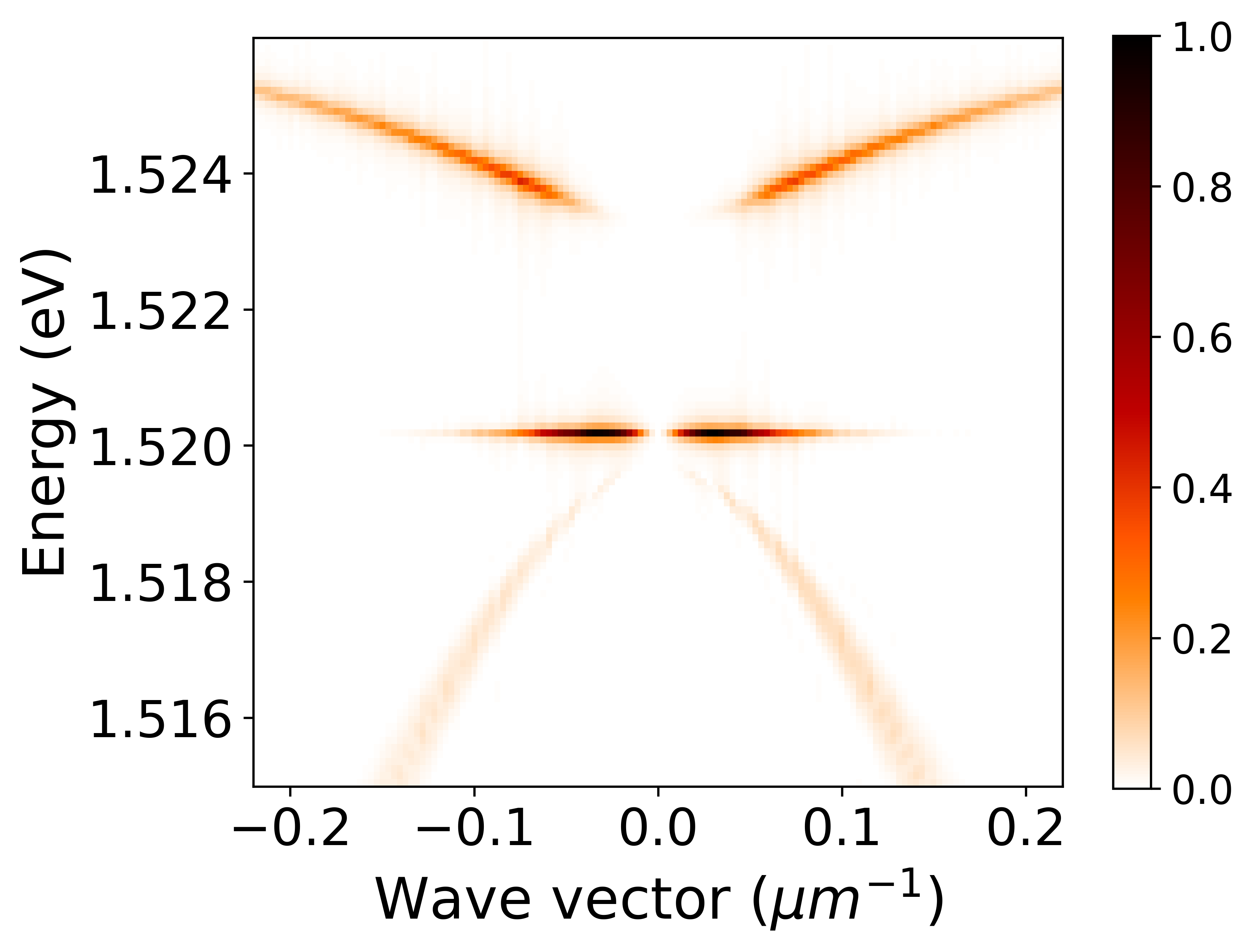}
    \put(-128,116){\textbf{d}}
  \end{subfigure}%
  \hspace{0.01\textwidth}%
  \begin{subfigure}{0.33\textwidth}
    \centering
    \includegraphics[width=\linewidth]{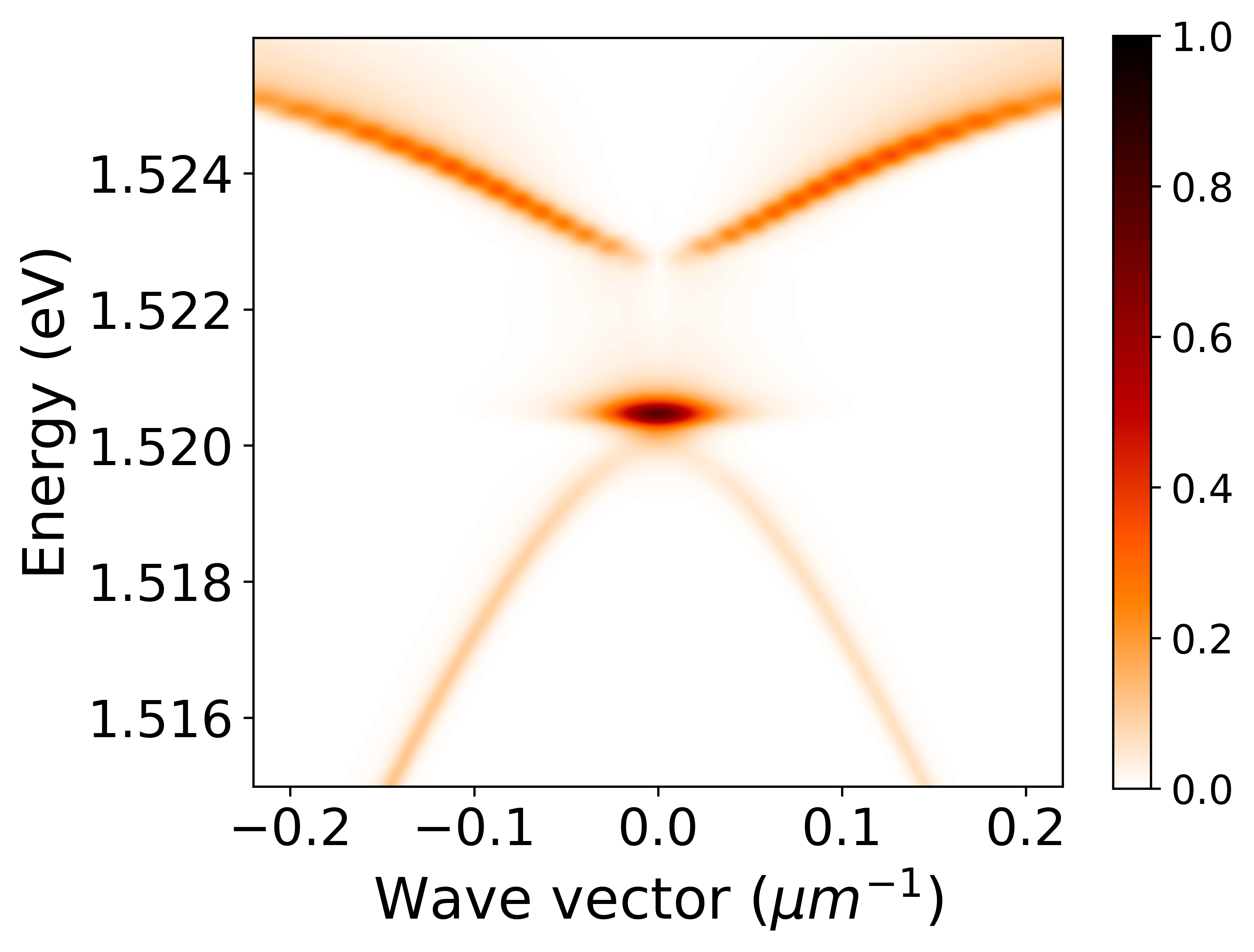}
    \put(-128,116){\textbf{e}}
  \end{subfigure}
  
  \caption{(a) Sketch of the etched grating in the GaAs/AlGaAs heterostructures and sample excitation/collection from the vertical direction. Measured polariton dispersion in gratings supporting (b) a lower BIC and an upper lossy mode, and (c) a lower lossy and an upper BIC; gap confined modes arising from either the BIC (b) or the lossy (c) branch, respectively, are clearly visible. The respective calculated excitonic fractions are 44 $\%$ and 47 $\%$.  
  (d,e) Out-of-equilibrium GPE simulations of polariton emission for the two situations measured in (b) and (c), respectively.}
  \label{fig:Fig1}
\end{figure*}

\textit{Sample design and fabrication}. 
The fabricated sample is composed of a waveguide core made of 12 GaAs quantum wells (QWs) 20 nm thick and 13 Al$_{0.4}$Ga$_{0.6}$As barriers grown on an Al$_{0.8}$Ga$_{0.2}$As cladding layer. On the same chip, we fabricated two different one-dimensional (1D) gratings etched in 170 nm deep grooves, allowing to engineer the polariton dispersion around normal incidence with the presence of either a BIC or a lossy photonic mode separated by an energy gap, as schematically shown in Fig.~\ref{fig:Fig1}a. Differently from previous works,\cite{Ardizzone2022PolaritonContinuum,Riminucci2022NanostructuredContinuum} here we etched deeper grooves in order to create a larger energy gap between upper and lower branches, whose BIC and lossy nature can be exchanged as a function of the air fraction\cite{Riminucci2022NanostructuredContinuum,Lu2020EngineeringPoints}. The QWs in the waveguide core support exciton resonances that are strongly coupled to the photonic modes, giving rise to polariton branches whose symmetry properties are derived from the ones of the corresponding photonic coupled modes \cite{Dang2022RealizationMetasurface}.
In fact, the details of the grating geometry, such as air fraction or number of grooves in the unit cell, determine whether the maximum energy point in the dispersion is a BIC (as seen, e.g., in Fig.~\ref{fig:Fig1}b) or a lossy state (e.g., Fig.~\ref{fig:Fig1}c). Numerical simulations of the optical response from these structures performed with a Maxwell solver are given in the Supplementary Information (SI). In particular, the grating designs were performed by using the Stanford Stratified Structure Solver (S4).\cite{Liu2012SStructures}.


\begin{figure*}
  \centering
  \begin{subfigure}{0.45\textwidth}
    \centering
    \includegraphics[width=\linewidth]{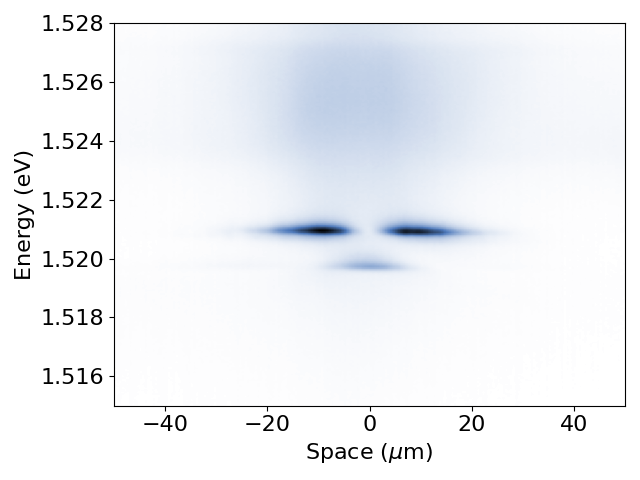}
    \put(-180,150){\textbf{a}}
  \end{subfigure}%
  \hspace{0.03\textwidth}%
  \begin{subfigure}{0.44\textwidth}
    \centering
    \includegraphics[width=\linewidth]{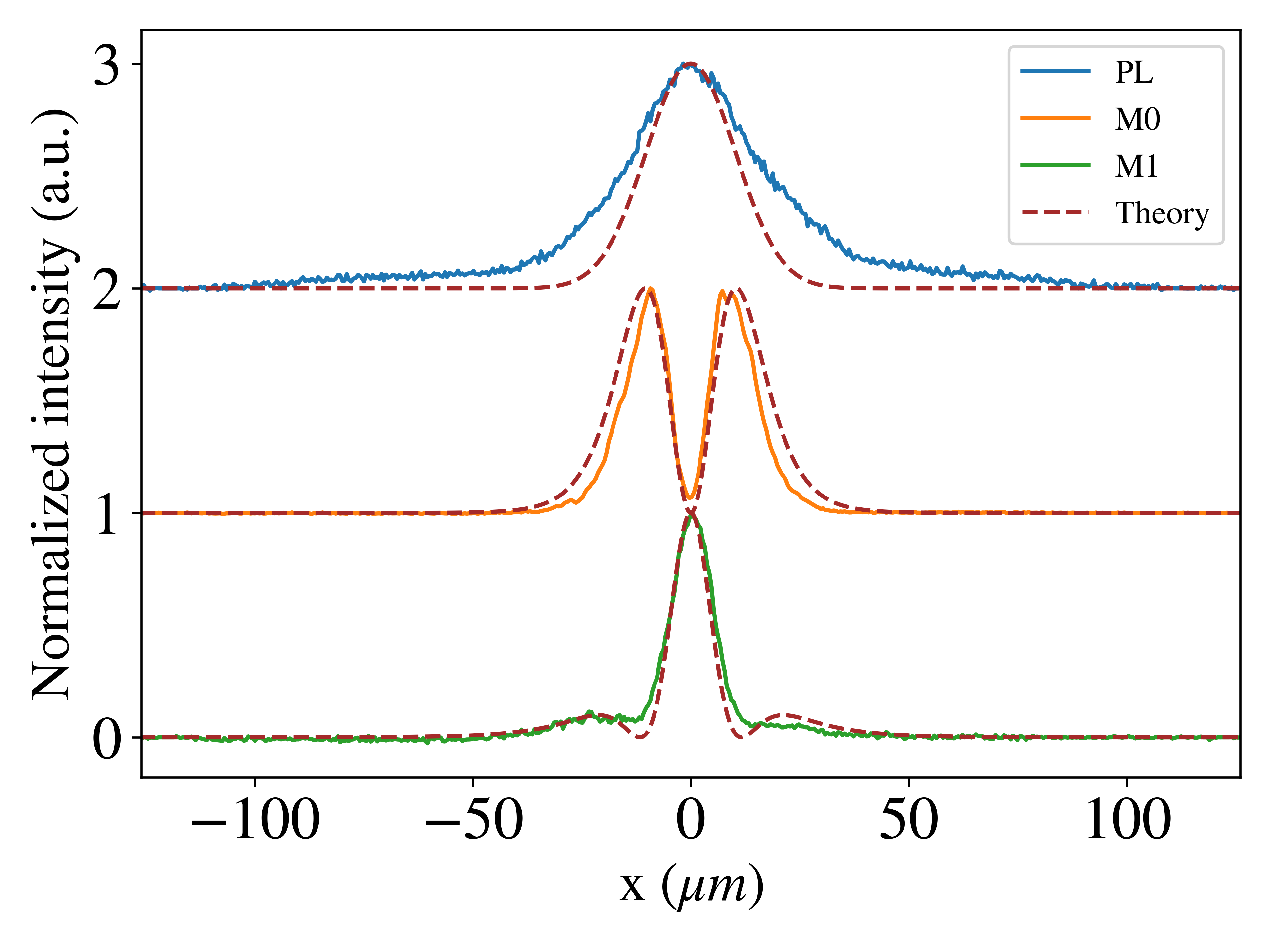}
    \put(-185,150){\textbf{b}}
  \end{subfigure}%
  \medskip
  
  \begin{subfigure}{0.45\textwidth}
    \centering
    \includegraphics[width=\linewidth]{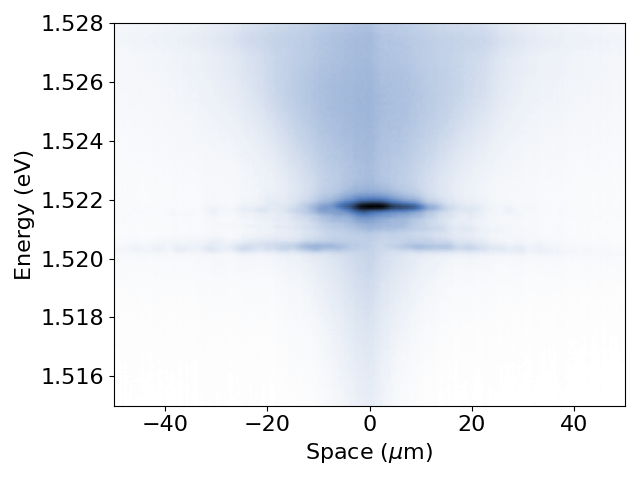}
    \put(-180,150){\textbf{c}}
  \end{subfigure}%
  \hspace{0.03\textwidth}%
  \begin{subfigure}{0.44\textwidth}
    \centering
    \includegraphics[width=\linewidth]{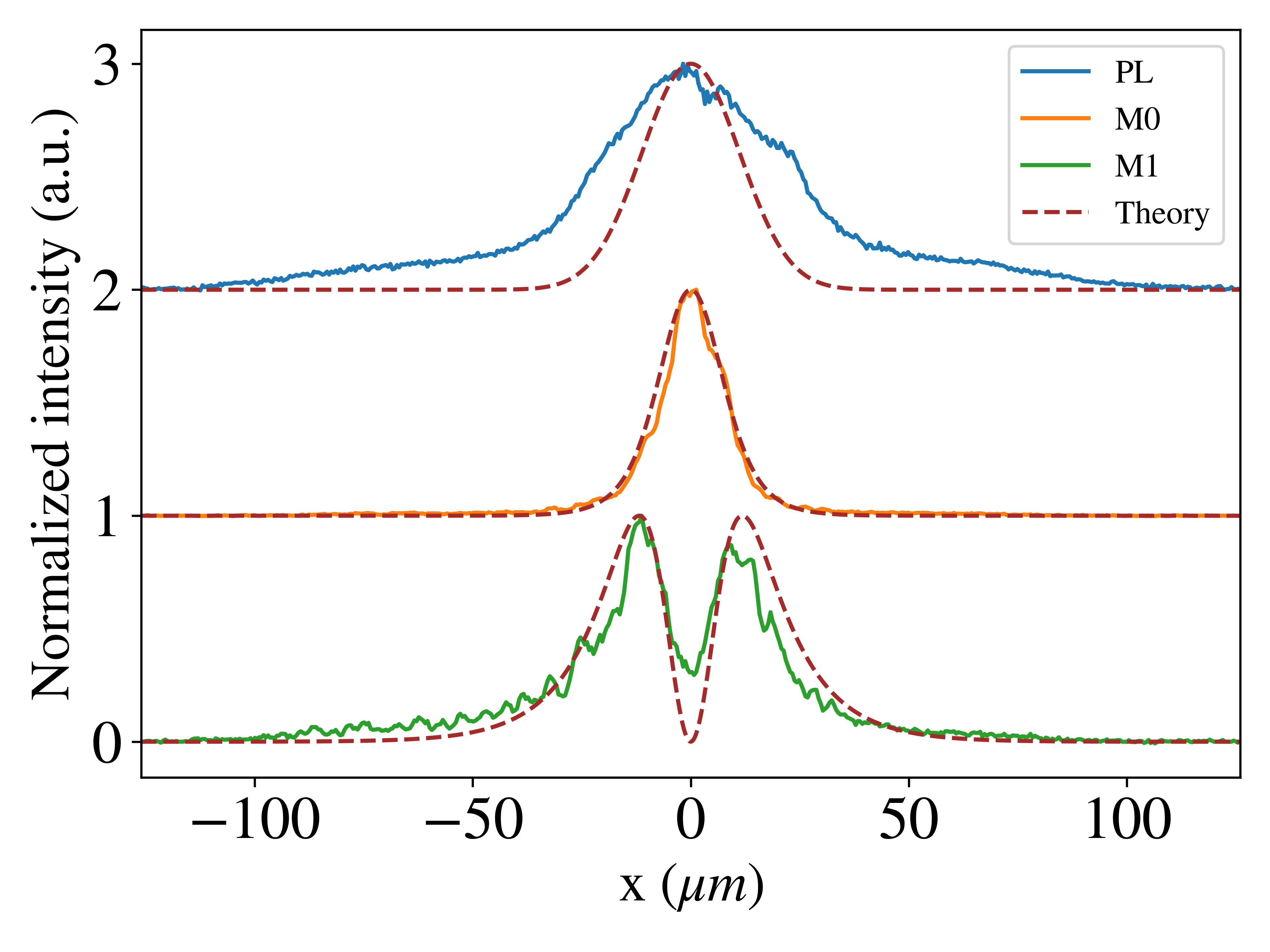}
    \put(-185,150){\textbf{d}}
  \end{subfigure}%
  
  \caption{Emission from spatially confined and quantized polariton condensate modes, evidencing a fundamental (M0) and first excited (M1) state within the potential well created by the exciting laser spot, arising from either a BIC (a,b) or a lossy (c,d) lower branch. In panels (b,d) the $x$-dependence of the measured luminescence is directly compared to the spatially resolved emission simulated through the out-of-equilibrium GPE formulation (brown dashed lines). The measured exciton emission (blue line, corresponding to the broad and weak emission at the energy 1527.5 meV) is also compared to the Gaussian confining potential assumed in the simulations.}
  \label{fig:Fig2}
\end{figure*}

\textit{Experiments}.
Measurements are performed in a cryostat at 4 K, and uncoupled excitons and polaritons are introduced by nonresonant pumping onto the grating with a laser tuned at 1.59 eV, with a 12 $\mu$m full width half maximum (FWHM) spot. All measurements were taken using a single mode continuous wave (cw) laser. However for threshold characterisation we used a pulsed laser with 80 MHz repetition rate and 100 fs pulse width to avoid spurious thermal effects that could arise at high pump powers under cw excitation. As the excitons diffuse in each QW plane, we expect the potential well to spatially broaden, as explained in the SI file. The photoluminescence (PL) signal (at around $\sim 1527.5$ meV) is collected through a 3 cm focal length air spaced doublet and the real/reciprocal space plane is reconstructed on the monochromator slits connected to a charged coupled device (CCD). At low excitation power, PL emission clearly evidences the dispersion of exciton-polariton branches, corresponding to extended modes along the grating. On the other hand, on increasing power, at the onset for polariton condensation, a localized mode appears in the energy gap, arising from the BIC negative effective mass branch below, as confirmed by its darkness at exactly $k=0$ $\mu m^{-1}$ in Fig.~\ref{fig:Fig1}b. The physical mechanism is similar to hole-confinement in semiconductors, therefore we will define it as a gap-confined mode henceforth.
Changing the grating geometry allows to swap the BIC with a radiative state around $k=0$ $\mu m^{-1}$\cite{Riminucci2022NanostructuredContinuum,Lu2020EngineeringPoints,maggiolini2022strongly}, as shown in Fig.~\ref{fig:Fig1}c. In fact, and this is one of the unexpected outcomes of this work, polariton condensation takes place in the gap-confined mode even if it is a lossy one and not a long-lived BIC, where it was previously shown \cite{Ardizzone2022PolaritonContinuum}. This indicates that the mode lifetime plays a secondary role in the condensation phenomenon itself, suggesting a crucial contribution of the negative polariton mass despite only along one direction of the saddle-like dispersion in the lower branch.  
\begin{figure}
  \begin{subfigure}{0.45\textwidth}
    \includegraphics[width=\linewidth]{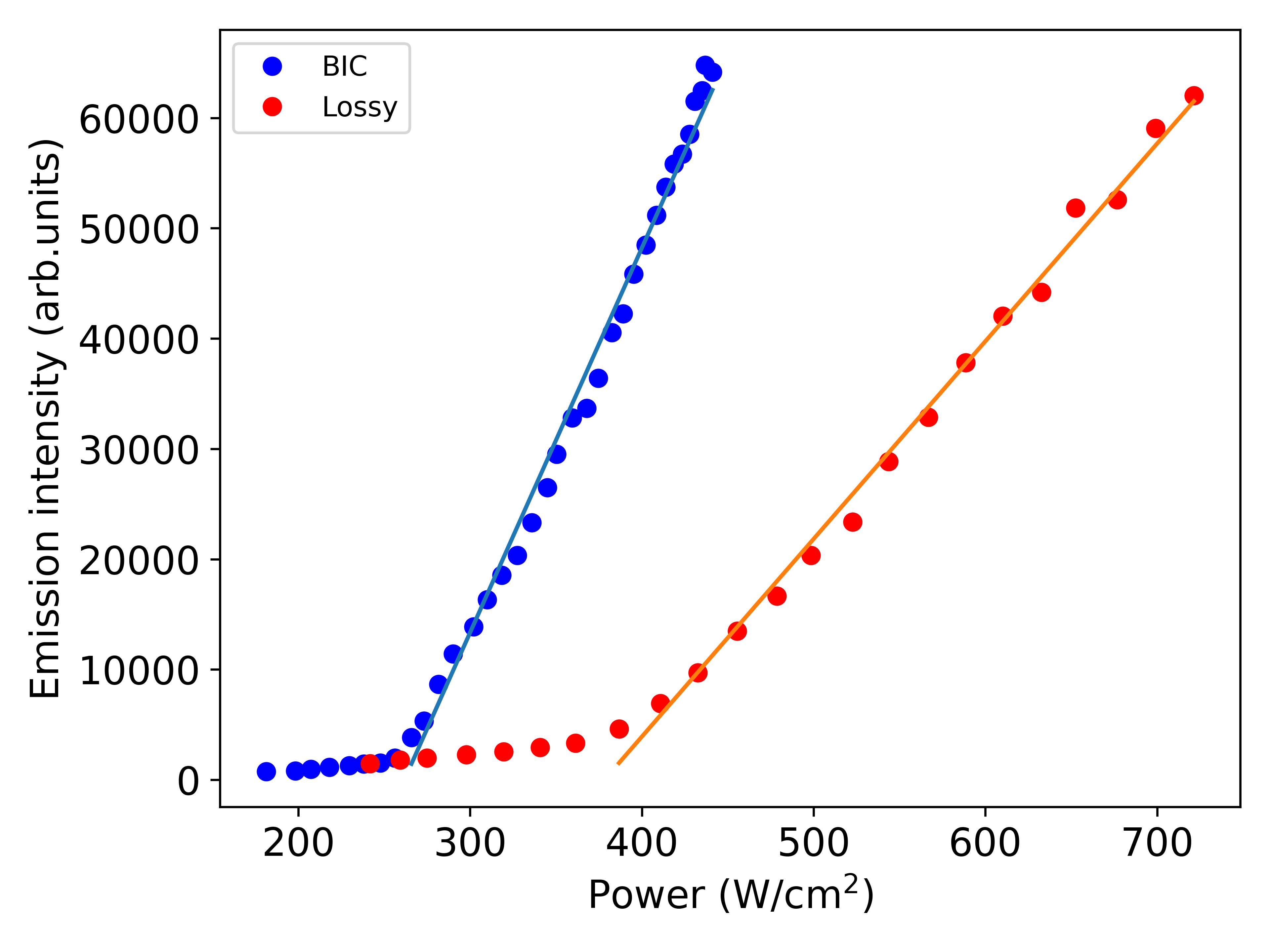}
    \caption{}
  \end{subfigure}%
  \medskip
  
  \begin{subfigure}{0.45\textwidth}
    \includegraphics[width=\linewidth]{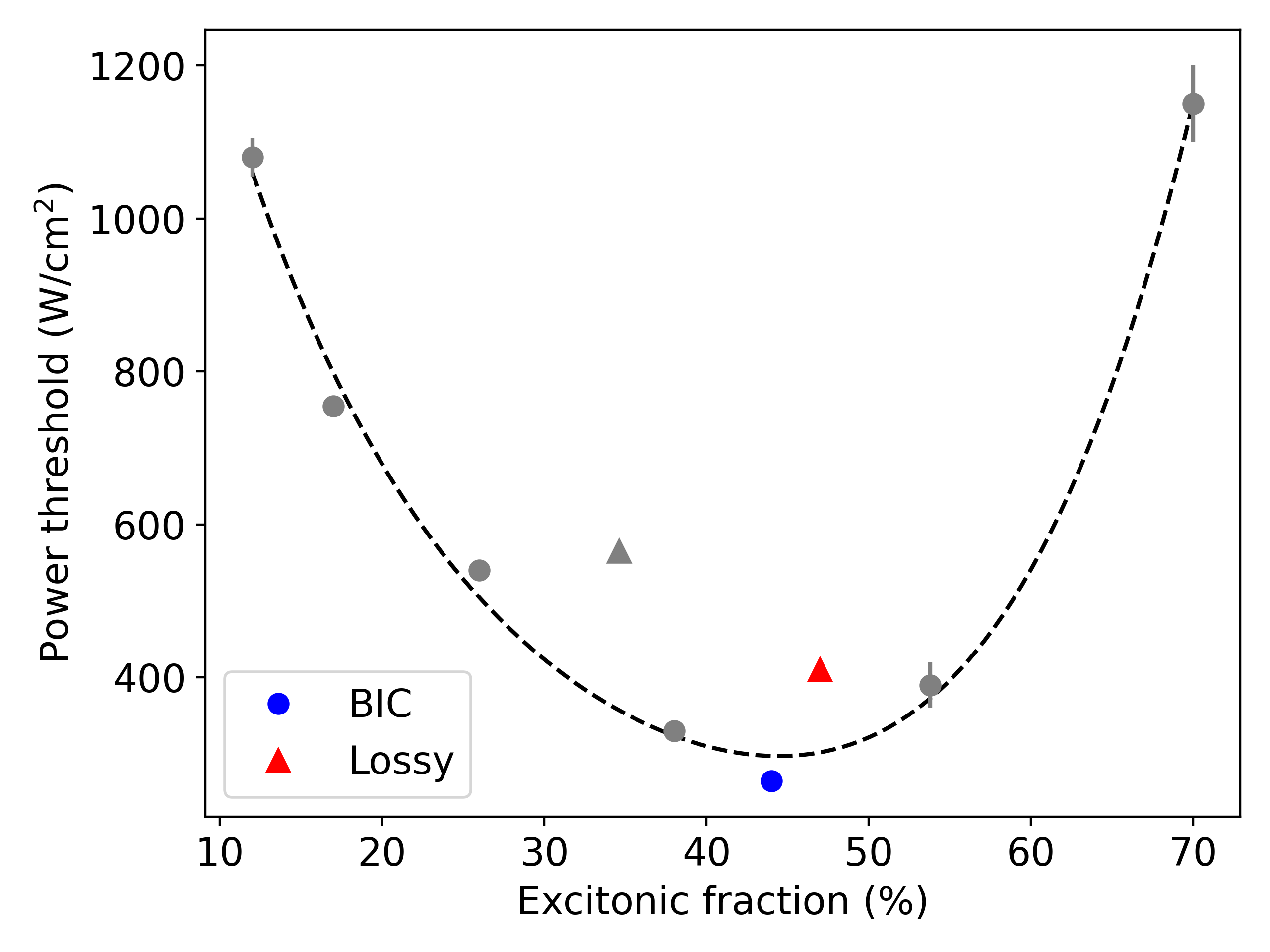}
    \caption{}
  \end{subfigure}%
  
  \caption{(a) Condensate emission intensity from the BIC (blue dots) and the lossy gap-confined state (red dots), respectively, plotted as a function of the incident laser power. (b) The circles and triangles represent the power thresholds measured on structures characterized by a condensate from a BIC and lossy, respectively, with different excitonic fractions. The 4th order polynomial fit through the condensation threshold from the BIC is represented by the dark dotted line. The condensate thresholds for the BIC and lossy shown in (a) are represented by the blue circle and red triangle, respectively.}
  \label{fig:Fig3}
\end{figure}

\textit{Theoretical modeling}.
In order to describe the condensation phenomenology shown from experiments, we resort to the formulation of an out-of-equilibrium Gross-Pitaevskii equation (GPE) under incoherent pumping \cite{Carusotto2013QuantumLight}, allowing to include the effects of exciton-photon dispersion and exciton-photon coupling in describing the dynamics of the system. Here we extend the previous formulation from Wouters and Carusotto \cite{Wouters2007ExcitationsPolaritons}, including the effects of multi-band photonic dispersions coupled to the QW exciton modes, wave-vector dependent loss rates within the single photonic band (distinguishing between BIC and lossy branches at $k=0$ $\mu m^{-1}$), and allowing for a confining potential created by the driven exciton population and described by a Gaussian function $P (x)= V_0 \exp\{-{x^2}/{2\sigma^2}\}$, where $2\sigma$ represents the spatial extent of the well, and $V_{0}$ its amplitude in meV. After solving for the multi-band polariton wavefunction (polariton state vector), emission can be calculated from the photonic components of each polariton eigenmode, after superimposing the positive and negative group velocity modes from each branch. An example of such emission calculations is shown in Fig.~\ref{fig:Fig1}d,e, where the gap-confined modes appear with the right symmetry inherited from the corresponding negative mass branch. Details of this extended formulation are summarized in the Supplementary Information, while a complete analysis of the model solutions will be reported elsewhere\footnote{D. Nigro and D. Gerace, in preparation}. 
We should notice that a 1D model is employed here, thus neglecting the full 2D dynamics induced, e.g., by the saddle-like dispersion of the lower branch, which does not seem crucial to correctly reproduce the experimental results (see Fig.~~\ref{fig:Fig1} and in the following).


\textit{Gap-confined modes and condensation threshold}.
By increasing the laser power, we observed different quantized states arising underneath the pumping spot. This effect was observed for both the BIC and lossy polariton modes, respectively. In Fig.~\ref{fig:Fig2} we show the PL emission resolved in real space, in which we clearly see a fundamental mode (M0) and first excited (M1) mode, for both the polariton condensates from the BIC (in Fig.~\ref{fig:Fig2}a,b) and the radiative negative mass branch (Fig.~\ref{fig:Fig2}c and d). The mode ordering is such that the M0 is higher in energy as compared to the M1 as it is appropriate for a negative mass confinement. The spatially dependent profiles of these modes for all the considered cases were also extracted from the out-of-equilibrium GPE solution, when assuming  a potential V$_0$ = 1.25 meV. These results are superimposed to the experimental plots in Fig.~\ref{fig:Fig2}b,d, showing an excellent agreement. The interpretation is straightforward: a local blueshift of the polariton dispersion is induced by the pump spot, which creates confined polariton states within the energy gap; the latter, in turn, has a photonic origin due to the dielectric profile modulation introduced from the grating. This local blue-shift creates an effective gaussian-like potential well in which negative mass excitations are confined, as anticipated before. For polaritons, this prevents scattering on the same modes outside the blue-shifted area. Furthermore, the negative mass is responsible for the presence of a trapping potential for polaritons, thus enhancing the relaxation towards the condensed mode as well\cite{Yoon2022EnhancedPotentials}. As long as the energy blue-shift induced by the pump spot does not overcome the energy gap, the available states within the well are gap-confined\cite{Ardizzone2022PolaritonContinuum}.\\

To clarify the role of polariton lifetime on the condensation mechanism, we performed power series measurements that allowed to observe the thresholds in the two cases considered in Fig.~\ref{fig:Fig1}, in which the BIC or the lossy mode are alternatively coming from the negative mass dispersion. The two cases have similar excitonic fraction associated with the energy state giving rise to the polariton condensate. Specifically, it is 44 $\%$ for the BIC and 47$\%$ for the lossy mode, and their respective thresholds are shown in Fig.\ref{fig:Fig3}a. In Fig.~\ref{fig:Fig3}b, we show the condensation thresholds for different excitonic fractions when the condensation originates from a BIC (circles) and lossy (triangles). The blue and red markers represents the threshold for the samples under consideration in Fig.\ref{fig:Fig3}a.
The fit through the BIC condensation thresholds shows that the considered excitonic fractions should not play an important role. Therefore, we can make a meaningful comparison between the two gratings. The condensation in the lossy state (red dots in Fig.~\ref{fig:Fig3}a) occurs at a higher power as compared to the one in the polariton BIC (blue dots in Fig.~\ref{fig:Fig3}a). The only meaningful difference between the two situations is related to the photonic losses of the relative condensed state. Moreover, as shown in Fig.~\ref{fig:Fig3}a, a different slope above threshold is observed, with an estimated ratio of 1.97 $\pm$ 0.02 from the experimental data. As described in Ref.~\cite{Wouters2007ExcitationsPolaritons}, the condensate density above threshold can be expressed as $|\Psi_{0}|^{2} \propto 1 /\gamma^{B,L}_{c}$, where $\gamma^{B,L}_{c}$ are the loss rates associated to the condensed mode that is either BIC ($B$) or Lossy ($L$). Although the BIC suppresses radiative losses, the condensate emits radiation in the far field at $k \neq 0$ $\mu m^{-1}$, as illustrated in Fig.\ref{fig:Fig1}a. This results in a radiative contribution, denoted as $\gamma_{rad}^{B}$. The corresponding loss rate from the lossy state is identified as $\gamma_{rad}^{L}$, whereas all losses that do not arise from photon emission in the far field are included in $\gamma_{nr}$. Based on the rate equations described in the SI, we can determine the ratio of power thresholds for the lossy and BIC, respectively 410 $W/cm^{2}$ and 265 $W/cm^{2}$, to be
\begin{equation}
    \frac{P_{th}^{L}}{P_{th}^{B}} = \frac{\gamma_{rad}^{L}+\gamma_{nr}}{\gamma_{rad}^{B}+\gamma_{nr}} = 1.55\pm 0.04
\end{equation}

Even though the BIC is suppressing the radiative losses at exactly normal incidence, the condensate is still susceptible to radiative emission at finite angles and nonradiative recombination in the material. 
As a result, we attribute the lower threshold and steeper curve to the suppressed perpendicular emission from the polariton BIC. In contrast to the BIC, the lossy state has additional radiative losses causing a shorter lifetime, which increases the threshold and decreases the observed slope above threshold.

\textit{Summary}.
We have shown the possibility to achieve Bose-Einstein condensation driven by a negative polariton mass dispersion in a waveguide embedded grating patterned on a multi-QW heterostructure, independently on whether the state in which the particles accumulate is a polariton BIC or a radiative state. The spatial quantization is demonstrated by the discretization of the energy levels induced by a confining potential created by the pumping spot itself, for both BIC and lossy modes. All these results have been successfully modeled by an extension of the 1D out-of-equilibrium GPE, including the effects of multi-band strong coupling, wave vector dependent losses, and effective confining potential in-plane. 
Finally, we studied the impact of photonic losses, observing changes in both the critical density and the dependence of the condensate population on the injection rate. This gives us the ability to estimate the ratio of losses between the two cases. From this analysis we can assert that while the BIC shows a lower threshold between the two cases, its longer lifetime is not the main factor to drive the condensation. Instead, our results show how the negative mass can be used to induce a trapping potential and drive the polariton thermalization in both cases towards the lower branch, thus triggering the condensation onset. 

\textbf{Acknowledgments}
We thank Paolo Cazzato for the
technical support.
The authors acknowledge the project PRIN-2017 ``Interacting Photons in Polariton Circuits INPhoPOL'' funded by the Ministry of University and Scientific Research (MUR), Grant No. 2017P9FJBS\_001. Work at the Molecular Foundry was supported by the Office of Science, Office of Basic Energy Sciences, of the U.S. Department of Energy under Contract No. DE-AC02-05CH11231. We acknowledge the project FISR—C.N.R. Tecnopolo di nanotecnologia e fotonica per la medicina di precisione—CUP B83B17000010001, and ``Progetto Tecnopolo per la Medicina di precisione, Deliberazione della Giunta Regionale'' Grant No. 2117. This research is partly funded by the Gordon and Betty Moore Foundations EPiQS Initiative, Grant No. GBMF9615 to L. N. Pfeiffer, and by the National Science Foundation MRSEC Grant No. DMR 1420541. We acknowledge support from the Project ``Hardware implementation of a polariton neural network for neuromorphic computing,'' the Joint Bilateral Agreement CNR-RFBR (Russian Foundation for Basic Reaserach), Triennal Programm 2021 -2023, the  Italian Ministry of Scientific Research (MUR) through the FISR2020–COVID, project ``Sensore elettro-ottico a guida d'onda basato sull'interazione luce-materia'' (WaveSense), contract no. FISR2020IP\_04324.


\bibliographystyle{unsrt}
\bibliography{sample}

\end{document}


\title{Supplementary Information to: \\
Bose-Einstein Condensates in Gap-Confined Exciton-Polariton States}

 \author{F. Riminucci}
 \affiliation{Molecular Foundry, Lawrence Berkeley National Laboratory, One Cyclotron Road, Berkeley, California 94720, USA}
 \author{A. Gianfrate}
 \affiliation{CNR Nanotec, Institute of Nanotechnology, via Monteroni, 73100, Lecce}
 \author{D. Nigro}
 \affiliation{Dipartimento di Fisica, Universit\`a di Pavia, via Bassi 6, Pavia (IT)}
 \author{V. Ardizzone}
 \affiliation{CNR Nanotec, Institute of Nanotechnology, via Monteroni, 73100, Lecce}
 \author{S. Dhuey}
 \affiliation{Molecular Foundry, Lawrence Berkeley National Laboratory, One Cyclotron Road, Berkeley, California 94720, USA}
 \author{L. Francaviglia}
 \affiliation{Molecular Foundry, Lawrence Berkeley National Laboratory, One Cyclotron Road, Berkeley, California 94720, USA}
 \author{K. Baldwin}
 \affiliation{PRISM, Princeton Institute for the Science and Technology of Materials, Princeton University, Princeton, New Jersey 08540, USA}
 \author{L. N. Pfeiffer}
 \affiliation{PRISM, Princeton Institute for the Science and Technology of Materials, Princeton University, Princeton, New Jersey 08540, USA}
 \author{D. Trypogeorgos}
 \affiliation{CNR Nanotec, Institute of Nanotechnology, via Monteroni, 73100, Lecce}
 \author{A. Schwartzberg}
 \affiliation{Molecular Foundry, Lawrence Berkeley National Laboratory, One Cyclotron Road, Berkeley, California 94720, USA}
 \author{D. Gerace}
 \affiliation{Dipartimento di Fisica, Universit\`a di Pavia, via Bassi 6, Pavia (IT)}
 \author{D. Sanvitto}
 \email{daniele.sanvitto@nanotec.cnr.it}
 \affiliation{CNR Nanotec, Institute of Nanotechnology, via Monteroni, 73100, Lecce}

\maketitle
\section{Grating design and fabrication}
The grating was fabricated by spinning ZEP520a 50$\%$ at 2000 rpm onto a 510 nm thick waveguide core made of GaAs/Al$_{0.4}$Ga$_{0.6}$As heterostructure. The resist was exposed using a Raith EBPG 5200 electron beam lithography tool, and developed in amyl-acetate for 1 minute. The etching was carried out in an Oxford ICP-Chlorine etcher, for a total etching depth of 170 nm. The resist was removed with dichloromethane, after which 10 nm of aluminum oxide were conformally deposited by means of atomic layer deposition. Fig. S1a shows a grating with a single groove of width w = 50 nm within a unit cell 242 nm wide. This grating creates a bound state in the continuum at the saddle point of the energy dispersion. The energy dispersion (Fig.~S1b) is simulated by using the Stanford Stratified Structure Solver (S4) as described in Ref.~[16]. The bright traces are related to the eigenmode dispersion, while the resonance linewidth is proportional to the wave-vector dependent radiative losses. The bound state in the continuum (BIC) is identified by a vanishing optical response at normal incidence. When a second groove is introduced within the unit cell (Fig.~S1c), the BIC at normal incidence switches to the upper branch below the exciton energy (see Fig.~S1d), as discussed in detail in Ref.~[23].

\begin{figure}[h]
\begin{center}
    \includegraphics[width=0.8\linewidth]{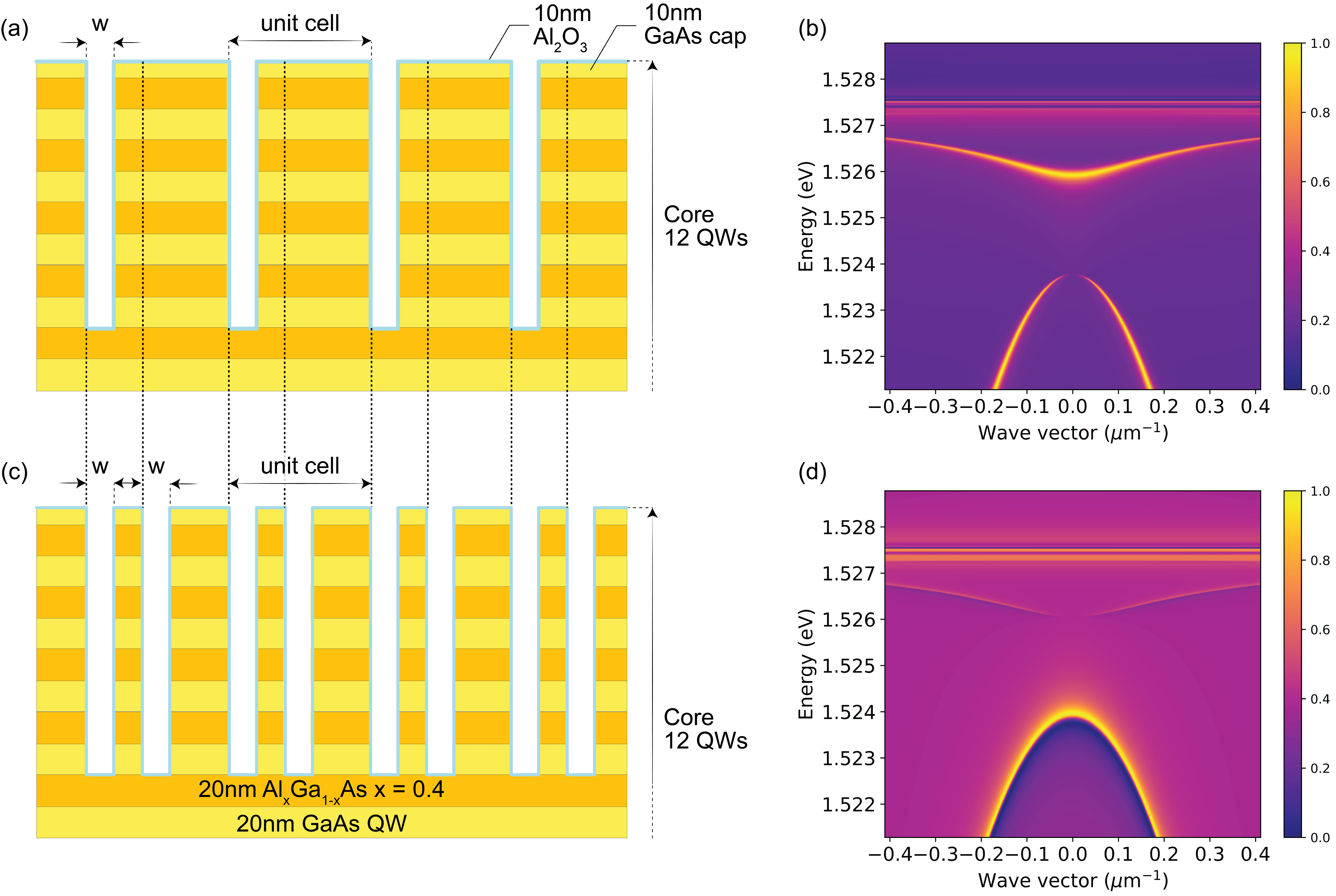}
     \caption{
     Sketches (a) and (c) show the waveguide core with two different geometries. The simulations of the respective energy dispersion are shown in (b), where the BIC occurs at the maximum of the negative mass dispersion branch, and in panel (d) which shows the BIC occurring on the upper branch and at zero wave vector.}
    \label{fig:sketch}
\end{center}
\end{figure}

\section{non-Hermitian Hamiltonian and rate equations model for condensation}
The Hamiltonian model used to calculate the polariton bands reads: 
\begin{equation}\label{eq:polariton_hamiltonian}
    H_0(x) = 
    \begin{pmatrix}
    E_x - i\hbar\gamma_x & \hbar\Omega_R & 0 & 0\\
    \hbar\Omega_R& \hbar\omega_0 + i \hbar v_g \partial_x-i\hbar\gamma_p & \hbar U - i\hbar \gamma_p &0\\
    0 &\hbar U - i\hbar \gamma_p & \hbar\omega_0 - i\hbar v_g \partial_x - i\hbar \gamma_p & \hbar \Omega_R\\
    0 & 0 & \hbar \Omega_R & E_x - i\hbar \gamma_x
    \end{pmatrix}
\end{equation}

This non-hermitian model accounts for two counter-propagating photonic modes with energy-momentum dispersion given by $\hbar\omega_0\pm i\hbar v_g \partial_x$, and two exciton flat-bands in momentum space located at energy $E_x$. Exciton-photon coupling is controlled by the Rabi energy $\hbar\Omega_R$. Due to the presence of the grating, counter-propagating modes are assumed to be diffractively coupled via the off-diagonal term $\hbar U$. The parameters $\hbar \gamma_x$ and $\hbar\gamma_p$ are phenomenological parameters used to account for the decay of the bare exciton and photon population, respectively, and they can be extracted from experimental data obtained in our samples far from the exciton-photon resonant condition. The polariton dispersion is then given by the real part of the eigevalues of Hamiltonian~\ref{eq:polariton_hamiltonian}, which can be directly compared to experimental spectra in energy-momentum space. \\
In fact, the values of the relevant parameters used in numerical simulations have been extracted by performing a best fit optimization, where we compared the polariton band structure obtained from the model in Eq.~\ref{eq:polariton_hamiltonian} for $\hbar \gamma_x=\hbar \gamma_p =0 \text{ meV}$, with the experimental data reported in the main text in Fig. 1b and 1c. In particular, by assuming an exciton energy $E_x=$1.5275 eV, the best fit optimization parameters obtained are as follows: $\hbar\Omega_R = 7\, \text{ meV}$; group velocity $v_g= 110 \,\mu \text{m/ps}$; photonic modes crossing $\hbar\omega_0= 1530.2 \,\text{ meV}$ ; diffractive coupling $\hbar U= 4.45 \,\text{ meV}$ and $\hbar U= -3.5 \,\text{ meV}$ for BIC and lossy, respectively. 
This behavior is in agreement with Ref. [19] of the main text, as we notice that the sign of the diffractive coupling term, $\hbar U$, determines whether the lower polariton branch is a BIC or a lossy eigenmode. Indeed, this is also confirmed from the results reported in Figs.~1d and 1e of the main text, where numerical simulations have been performed by assuming $\hbar U=4.45\,\text{ meV}$ and $\hbar U=-3.5\,\text{ meV}$ respectively, while keeping all the other parameters set to the values reported above.\\

{In fact, theoretical results reported in Figs.~1 and 2 in the main text have been obtained by numerically simulating the time-evolution of a four-components wavefunction, i.e. $\overrightarrow{\psi}(x,t)=(X_+(x,t),A_+(x,t),A_-(x,t),X_-(x,t))^T$ , with $A_{\sigma}$ and $X_{\sigma}$ denoting the photonic and excitonic fields associated to right- ($\sigma=+$) and left-moving ($\sigma=-$) excitations, respectively. In particular, our theoretical analysis has been performed by means of the following set of dynamical equations:
\begin{equation}\label{eq:dynamical_system}
\left\{
\begin{split}
\frac{d}{dt}n(x,t) & =P(x)-\gamma_x\,n(x,t)-  n(x,t) \sum_{l=1}^{4}\tilde{g}_{l}\vert\psi_{l}(x,t)\vert^2\\
\frac{d}{dt}\psi_l(x,t)&= \sum_{m=1}^4\tilde{H}_{l,m}(x)\psi_{m}(x,t)+\frac{1}{2}n(x,t)\,\tilde{g}_{l}\,\psi_{l}(x,t)\\
\end{split}\right. 
\end{equation} 
The dynamical system accounts for the driven-dissipative dynamics of a high-energy excitons reservoir density, defined by $n(x,t)$, in the presence of a nonlinear coupling to the relevant light-matter modes of the heterostructure, described by the components $\psi_{j}(x,t)$ of the vector $\overrightarrow{\psi}(x,t)$. Such a coupling is controlled by the phenomenological scattering parameters, $\tilde{g}_l$, which in the present case have been considered as having the same magnitude for the two photonic branches and the two excitonic bands for simplicity. We notice that the physical conclusions do not depend on this choice. To account for the spatially dependent pump profile, the Hamiltonian model $\tilde{H}(x)$ appearing in Eq.~\eqref{eq:dynamical_system} is obtained by adding a term proportional to the Gaussian pump-spot $P(x)$ (see main text) to each diagonal entry of $H_0(x)$. Indeed, these extra terms, which ultimately account for local changes introduced by the external driving in refractive indices affecting the photon dynamics as well as the local blue-shift induced on the flat exciton dispersion, are the ones responsible for the appearence of the gap-confined discrete polariton states.\\
In our protocol, an initial configuration corresponding to an empty reservoir and a weakly populated randomly initialized $\overrightarrow{\psi}$ is evolved in time according to Eq.~\eqref{eq:dynamical_system}. The overall time evolution is then used to extract the emission profiles reported in Figs.~1e and 1f of the main text. In particular, the results shown in the manuscript correspond to the normalized spectral density associated to the superposition of the photonic components $A_{\pm}(x,t)$, defined as  
\begin{equation}
I(k,E)= \vert A_{+}(k,E)+ A_{-}(k,E)\vert^2 \, ,
\end{equation}  
with $A_{\pm}(k,E)$ corresponding to the space-time Fourier transform of the components $A_{\pm}(x,t)$ over the entire space and over a finite time interval, $\Delta T$. Specifically, data reported in the main text correspond to $\Delta T=60$ ps.\\
Above condensation threshold, light emission becomes peaked at the energy of the gap-confined states, and polariton bands are no longer visible. In particular, results reported Figs.~2b and 2d in the main text provide an example of above threshold real-space emission profiles calculated as 
\begin{equation}
I_n(x)= \vert A_{+,n}(x)+ A_{-,n}(x)\vert^2 \, ,
\end{equation}
when only the modes M0 ($n=0$) and M1 ($n=1$) are macroscopically populated. Further details and an extensive analysis of the theoretical framework used in the present work are provided in Ref.~[26].
}

\section{Simplified rate equation model for condensation threshold}
To estimate the differences in condensation thresholds observed in the experiment for the two cases considered, we use an oversimplified model of two coupled rate equations, not related to the previously derived multi-band theory. In fact, assuming only a condensed mode with density $N_c$, and no spatial dependece of the pump rate, $P$, the system can simply be modeled using the following set of two coupled equations
\begin{equation}
    \frac{\partial N_R}{\partial t} = -\gamma_x N_R -gN_c N_R + P 
\end{equation}
\begin{equation}
    \frac{\partial N_c}{\partial t} = -\gamma_c N_c + g N_c N_R 
\end{equation}
where $N_R$ represents the density of the excitonic reservoir, 
$\gamma_x$ and $\gamma_c$ represent the reservoir and condensate losses, respectively. The coupling between the two levels is described by the coupling factor $g$. 
As the dynamics is determined by the slow exciton decay, we can assume a stationary state for which the condensate density $N_c^{B,L}$ for the BIC (B) and lossy (L) cases can be written as 
\begin{equation}
N_{c}^{B,L} = \frac{P}{\gamma_{c}^{B,L}}-\frac{\gamma_x}{g} = \frac{P-P_{th}}{\gamma_{c}^{B,L}} = \frac{P-P_{th}}{\gamma_{rad}^{B,L} + \gamma_{nr}}
\end{equation}
where $\gamma_{rad}^{B,L}$ is the loss rate from the emission in the farfield. All other losses are included in the term $\gamma_{nr}$ such that $\gamma_{c}^{B,L}=\gamma_{nr}+\gamma_{rad}^{B,L}$.

\section{Condensate from a more photonic lossy state}
\begin{figure}
\begin{center}
    \includegraphics[width=0.6\linewidth]{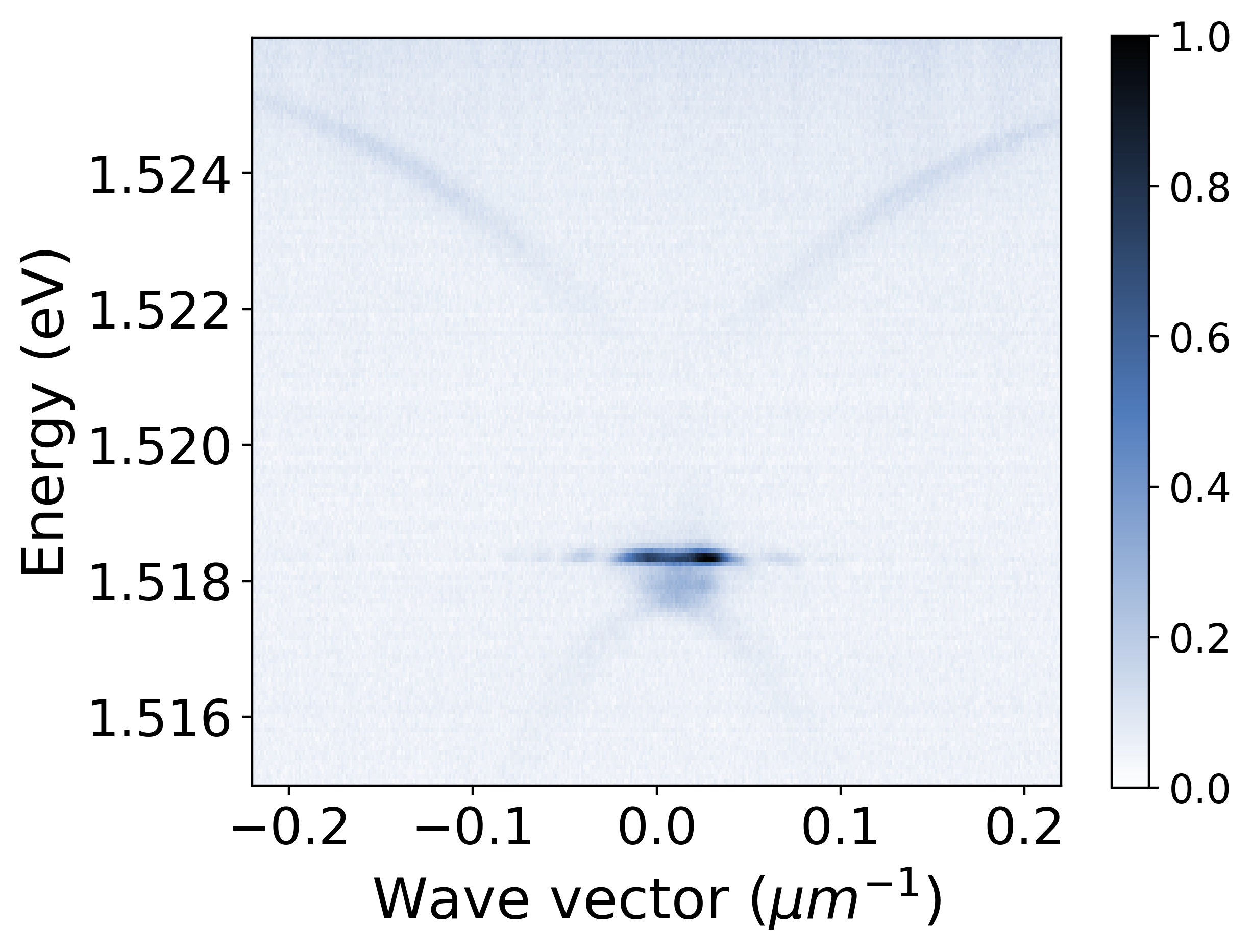}
     \caption{Exciton-Polariton dispersion above condensation threshold for a more photonic sample.}
    \label{fig:LowExBIC}
\end{center}
\end{figure}

In this section, we show that in the sample with a lossy state in the negative mass branch, the condensation still occurs from such point even when both the BIC and the lossy states are more photonic. This allows to exclude the possibility that condensation does not occur in the BIC in the sample considered in the main text due to higher losses resulting from high excitonic fractions. Fig.\ref{fig:LowExBIC} depicts the condensation originating from the lossy, with an excitonic fraction of 34.6$\%$.

